\RequirePackage{fix-cm}
\documentclass[twocolumn,epjc3]{svjour3}  
\smartqed  
\RequirePackage{graphicx}
\RequirePackage{amssymb}
\usepackage{gensymb}
\usepackage{dblfloatfix} 
\usepackage[caption=false]{subfig}
\usepackage{changes}
\usepackage[numbers]{natbib}
\usepackage[switch]{lineno}
\usepackage{overpic}
\usepackage{orcidlink}
\usepackage[title]{appendix}
\usepackage{algorithm}
\usepackage{algpseudocode}
\usepackage{enumitem}

\usepackage{xspace} 
\newcommand{\hess}{H.E.S.S.\xspace}
\usepackage{array}
\algnewcommand\algorithmicnot{\textbf{not}}
\algdef{SE}[IF]{IfNot}{EndIf}[1]{\algorithmicif\ \algorithmicnot\ #1\ \algorithmicthen}{\algorithmicend\ \algorithmicif}%

\newcolumntype{P}[1]{>{\centering\arraybackslash}p{#1}}
\journalname{Eur. Phys. J. C}
\begin{document}

\title{Background rejection using image residuals from large telescopes in imaging atmospheric Cherenkov telescope arrays}

\titlerunning{Image residuals for background rejection in IACTs}  

\author{L.~Olivera-Nieto~\orcidlink{0000-0002-9105-0518}\thanksref{e1,addr1}
        \and
        H.~X.~Ren~\orcidlink{0000-0003-0221-2560}\thanksref{addr1}
        \and
        A.~M.~W.~Mitchell~\orcidlink{0000-0003-3631-5648}\thanksref{addr2} 
        \and 
        V.~Marandon~\orcidlink{0000-0001-9077-4058}\thanksref{addr1}
        \and
        J.~A.~Hinton~\orcidlink{0000-0002-1031-7760}\thanksref{addr1} }

\thankstext{e1}{e-mail: Laura.Olivera-Nieto@mpi-hd.mpg.de}


\institute{Max-Planck-Institut für Kernphysik, P.O. Box 103980, D 69029, Heidelberg, Germany \label{addr1}
           \and
           Friedrich-Alexander-Universit\"at Erlangen-N\"urnberg, Erlangen Centre for Astroparticle Physics, 91058 Erlangen, Germany \label{addr2}
}

\date{Received: 10 August 2022 / Accepted: 23 November 2022}

\maketitle

\begin{abstract}
Identification of Cherenkov light generated by muons has been suggested as a promising way to dramatically improve the background rejection power of Imaging Atmospheric Cherenkov Telescope (IACT) arrays at high energies. However, muon identification remains a challenging task, for which efficient algorithms are still being developed. We present an approach in which, rather than identifying Cherenkov light from muons, we simply consider the presence of Cherenkov light other than the main shower image in IACTs with large mirror area. We show that in the case of the H.E.S.S. array of five telescopes this approach results in background rejection improvements at all energies above 1 TeV. In particular, the rejection power can be improved by a factor $\sim3-4$ at energies above 20 TeV while keeping $\sim90\%$ of the original gamma-ray efficiency.
\end{abstract}

\section{Introduction}
\label{sec:intro}

\begin{figure*}[h!]
    \centering
	\includegraphics[width=0.9\textwidth]{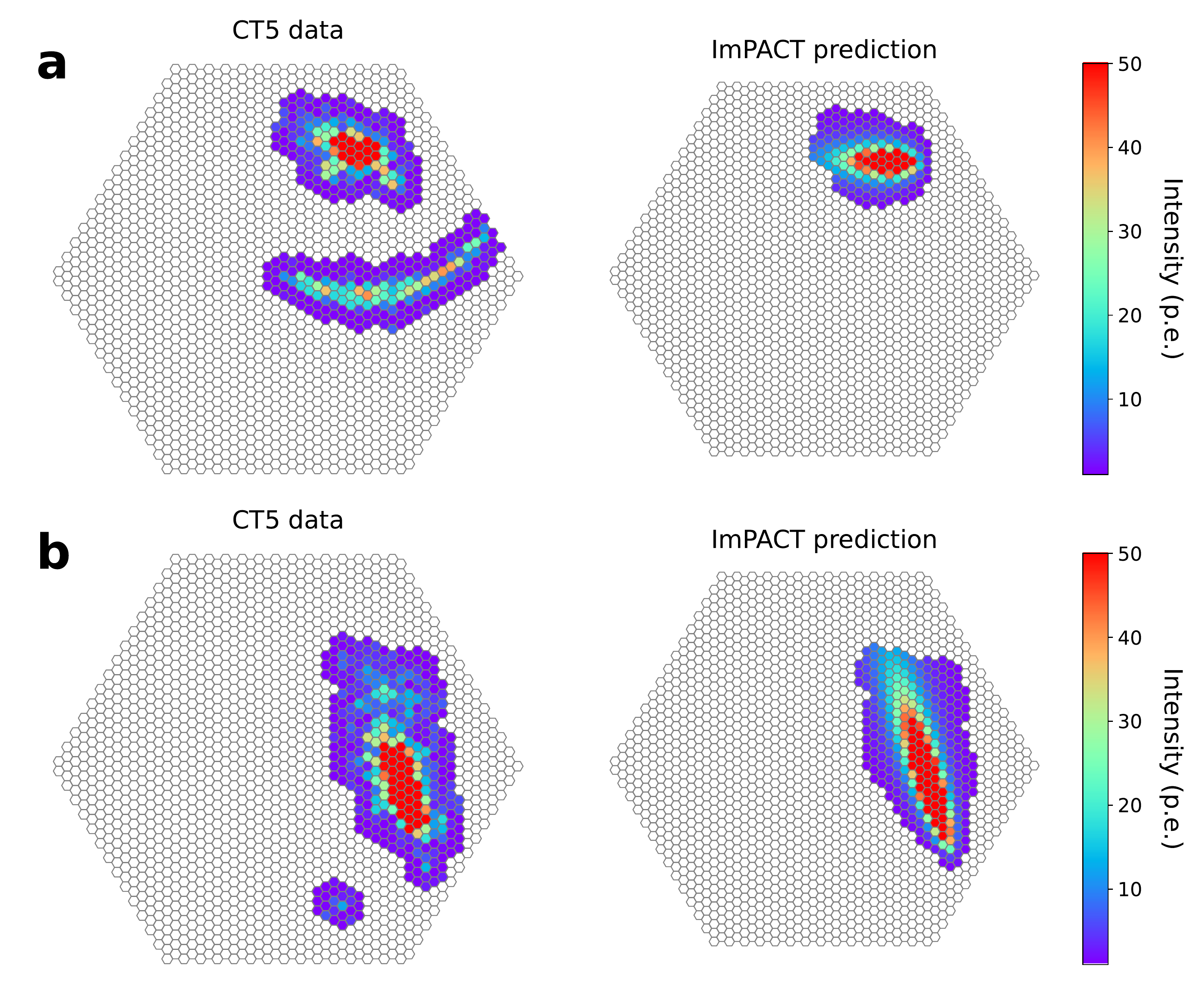}
	\caption{Large telescope images of rejected events that would be labelled as gamma-ray candidates in the small telescope reconstruction. In both cases, the right panel shows the ImPACT prediction associated with that event, whereas the left panel shows the actual recorded event image. The event image shown in the panel labelled as "a" contains a clear muon arc, whereas the additional feature in the image in panel "b" is much smaller and has a simpler shape. However, the maximum pixel intensity in this feature is more than double what is expected from NSB noise. Note that the colorbar has been restricted so that the fainter features are visible, since the main shower is much brighter in both cases.}
	\label{fig:example_off}       
\end{figure*}


The Earth's atmosphere is opaque to very-high-energy (VHE) radiation. Incoming gamma-ray photons lose their energy by initiating electromagnetic particle cascades in the atmosphere. The particles in these cascades are highly relativistic and cause the production of Cherenkov light, which is in turn collected by the telescope dishes of Imaging Atmospheric Cherenkov Telescopes (IACTs). Given the huge rates of background cosmic-rays, a factor $10^4$ greater than those of gamma-rays~\cite{background-factor}, differentiating between hadronic and electromagnetic cascades is a task critical to gamma-ray astronomy.

IACTs have superior rejection power to other ground-based arrays in the domain around 1 TeV, exploiting primarily the differences in shower width and substructure between electromagnetic and hadronic showers~\cite{stefanBDT}. At higher energies, of around tens of TeVs the rejection power worsens. High-energy events are often not fully contained in the camera, which makes the determination of the image parameters difficult. Additionally, this parameterisation becomes dominated by the very bright central component of high-energy showers, with additional faint emission that would indicate a hadronic origin being overlooked.

Large numbers of muons, primarily resulting from the decay of charged pions, are usually produced in the hadronic cascades initiated by cosmic-rays. The potential of muons as a tool to separate between these two classes of showers has long been recognized (see, for example~\cite{Gaisser}). Recently, \cite{ourpaper} showed that for very large ($\gtrsim$20~m mirror diameter) telescopes efficient identification of muon light can potentially lead to background rejection levels up to $10^{-5}$ at energies above 10~TeV.

However, muon identification is not a trivial task. Muons that pass directly through the telescope dish produce a distinguishable ring-like image in the telescope camera. But once the ground impact distance increases beyond some tens of meters only a fraction of the ring arc is detected (see top panel of Figure~\ref{fig:example_off}). Therefore muons which arrive far from the telescopes can easily be confused for low-energy shower images, or optical night-sky-background (NSB) noise. Alternative approaches, like making use of the signature of arrival time of muon light are promising, yet even then the task of muon identification remains a challenge~\cite{timing}.

Nevertheless, given the magnitude of the achievable background rejection power shown in~\cite{ourpaper}, less-than efficient muon identification could still result in a significant improvement to the current background rejection power of IACTs. We present an approach in which, rather than being concerned with whether a recorded event contains muon light or not, we simply consider the presence of light other than the main shower image (see bottom panel of Figure~\ref{fig:example_off}). This, of course, has the downside of being less precise in the sense that light from particles other than muons - electrons, for example - may be used to reject events. Additionally, unusually high NSB noise could lead to an event being rejected. However, it results in a much simpler implementation which, as shown below, still leads to improved background rejection power. 

For this task we used data and simulations from the telescopes in the High Energy Stereoscopic System (\hess)~\cite{hess2}. The \hess experiment is comprised of a total of five telescopes: four with a dish of 12~m diameter referred to as CT1-4 and a central one, CT5, with a dish of 28~m diameter. The event reconstruction is referred to as \textit{stereo} when only the small CT1-4 telescopes are used, and \textit{hybrid} when data from the entire array is used. In this work we present an alternative approach in which only the data of CT1-4 is used for the event reconstruction, that is, \textit{stereo} mode, but the data of CT5 is used for an extra step of background rejection.

\section{Background rejection with image residuals: \textit{ABRIR}} 
\label{sec:description}

ABRIR (\textit{Algorithm for Background Rejection using Image Residuals}) is a background rejection algorithm which draws additional information from the event image taken by a large telescope, such as CT5 for the case of the \hess array. This algorithm is applied after the usual \hess \textit{stereo} reconstruction, which includes an initial step of background rejection based on Boosted Decision Trees (BDTs)~\cite{stefanBDT}. In particular, we use two sets of initial cuts: the \hess \textit{standard} selection cuts (see Section 4.2 of~\cite{stefanBDT}), optimized for a Crab Nebula-like source and the so-called \textit{hard} cuts, optimized for a faint hard source. The main difference between both sets of cuts is that while the \textit{standard} cuts only require images to have a total intensity larger than 60 photoelectrons (p.e.), the \textit{hard} cuts require a minimum of 200 p.e.. This translates into a higher energy threshold but also higher quality events and more precise reconstructed parameters, such as core location or energy. Additionally the threshold of the BDT parameter~\cite{stefanBDT} is different, with 0.84 for the \textit{standard} case and 0.8 for the \textit{hard} cut. We will denote the gamma-ray and background efficiency of this first cut as $\eta_{BDT, \gamma}$ and $\eta_{BDT, B}$ respectively. We will apply ABRIR only to the events that survive this initial cut. Only at this step is the data from CT5 used, and it is exclusively CT5 images being considered by the algorithm due to the advantaged muon detection capabilities of large telescopes. Note that the total gamma-ray and background efficiency ($\eta_{tot, \gamma}$ and $\eta_{tot, B}$) needs to be computed as the product of that of the initial cut and that of ABRIR, so:
$$
\eta_{tot, \gamma} = \eta_{\gamma}  \cdot \eta_{BDT, \gamma}
$$
$$
\eta_{tot, B} = \eta_{B}  \cdot \eta_{BDT, B} 
$$
where $\eta_{\gamma}$ and $\eta_{B}$ are the gamma-ray and background efficiency of ABRIR (see Section~\ref{sec:peformance}).

\subsection{The ImPACT templates} 
\label{subsec:ImPACT}
\sloppy
In order to identify light as not a part of the main shower, we need to identify the main shower itself. We do this with the help of the \textit{Image Pixel-wise fit for Atmospheric Cherenkov Telescopes} (ImPACT) algorithm~\cite{impact}, which is routinely used by the H.E.S.S experiment. ImPACT is a gamma-ray event reconstruction algorithm that is based on the likelihood fitting of camera pixel amplitudes to an expected image template. These templates are built with a full Monte Carlo gamma-ray air shower simulation, followed by ray-tracing of the telescope optics and simulation of the instrument electronics~\cite{sim-telarray, corsika}. This fit is computationally expensive, so it is only performed after a first round of background rejection is applied. For each event this process results in an image of what the most similar gamma-ray event would look like in each of the telescopes. We use this prediction to mask the main component of CT5 image in order to better identify residual features. Note that the CT5 image is not used in this likelihood fit, with the CT5 template being derived only from the best-fit parameters of the CT1-4 fit.

\subsection{The rejection criteria} 
\label{subsec:algorithm}
After the main shower is masked, the remaining residual image is searched for clusters of pixels (N). A cluster is only considered if it is comprised of more than three neighboring pixels. For each of the clusters we compute the maximum intensity $I_{\mathrm{max, N}}$, the total intensity $I_{\mathrm{tot, N}}$ and the distance to the main shower $d_{\mathrm{N}}$. In the next step, two conditions are checked against each of the clusters:
\begin{enumerate}[label=(C\arabic*),wide=0pt, leftmargin=*]
    \item $I_{\mathrm{tot, N}} \cdot d_{\mathrm{N}}^2 > I_{\mathrm{tot}} \cdot d^2 $
    \item $I_{\mathrm{max, N}} > I_{\mathrm{max}}$
\end{enumerate}
where the threshold values used here for $I_{\mathrm{max}}$ and $I_{\mathrm{tot}} \cdot d^2$ are 9 p.e. and 2 p.e $\cdot$ pixel$^2$ respectively. These values were selected to maintain a gamma-ray efficiency of around 90\%, which in turn results in the background rejection performance described in Section~\ref{sec:peformance}. A different requirement on the gamma-ray efficiency would naturally impact the background rejection power, with better rejection achievable if a higher fraction of gamma-rays is lost. Additionally, these values are specific to the current state of the central telescope in the \hess array and should be adjusted as appropriate for application to other telescopes or cameras. 

The first condition (C1) selects clusters by their total intensity with a penalization on proximity to the main shower. This is done to reject clusters created solely by a mismatch between the outer row of pixels of the main shower image and the predicted template. Such clusters might have a high total intensity, but they are also very close to the main shower, which means the product of intensity and distance will fall below the required threshold. Additionally, this cut is always survived by relatively bright clusters that are far from the shower, and thus have a high probability of not originating from a mismatch between the data and the template. The second condition (C2) aims to reject clusters resulting from uncleaned noise from the NSB. It does so by requiring the maximum pixel intensity of a given cluster to be above a threshold defined by the 5$\sigma$ level of the pedestal width of a typical run. Note that this specific value depends on the image cleaning method used, as well as on the camera performance and expected NSB level.

\begin{figure}
    \centering
	\includegraphics[width=0.5\textwidth]{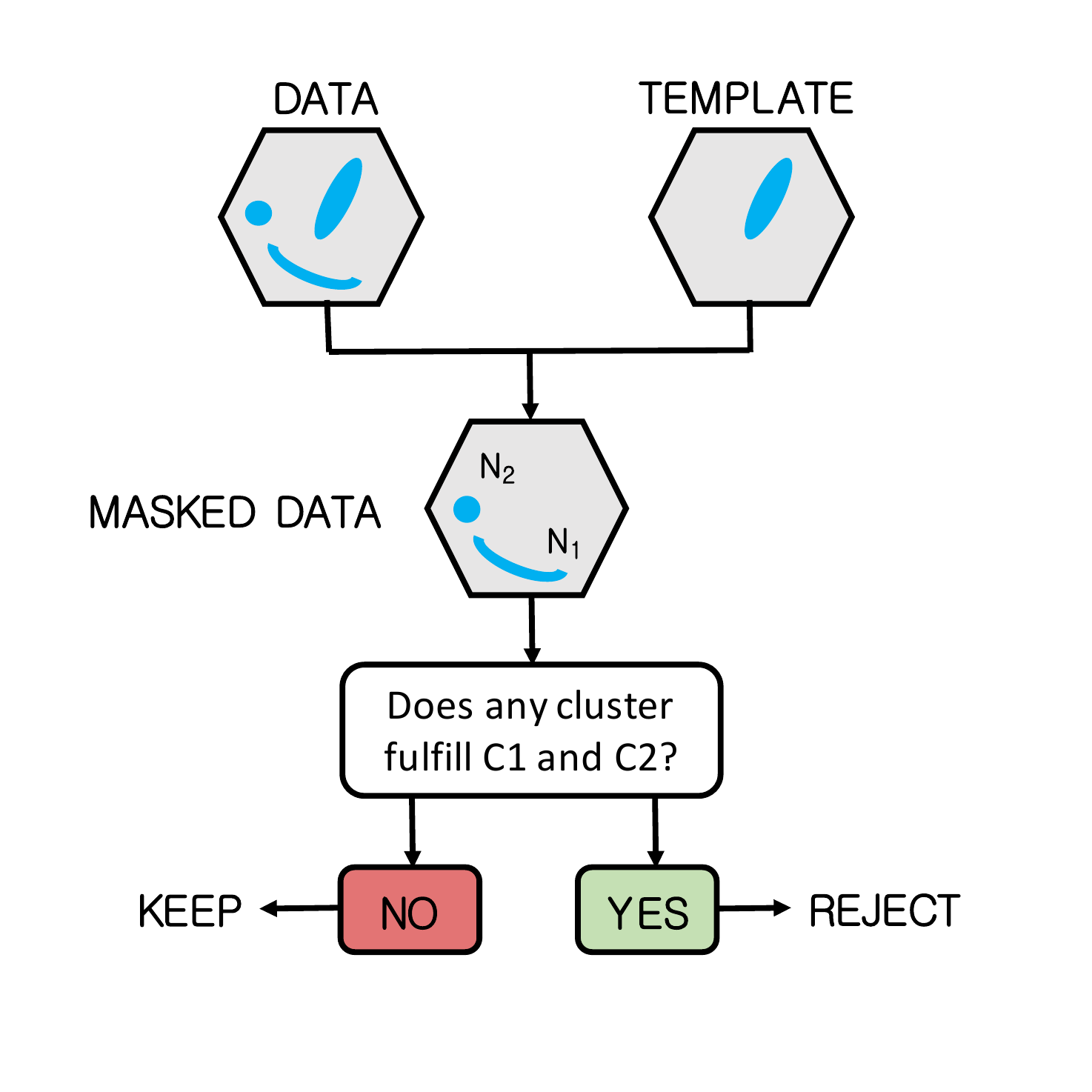}
	\caption{Schematic diagram representing the algorithm structure.}
	\label{fig:diagram}       
\end{figure}

Events that fail either of the two conditions are kept, as well as events with too small or no clusters outside of the main image. A schematic view of the algorithm is shown in Figure~\ref{fig:diagram}. Additionally, events for which no light in the large telescope was predicted by the template, but something is seen are also marked as rejected, as well as the opposite case, events in which an shower image was predicted, but all pixels in the data are below a minimal threshold of 1 photoelectron.

    



\section{Performance}
\label{sec:peformance}
\begin{figure*}
    \centering
	\includegraphics[width=0.95\textwidth]{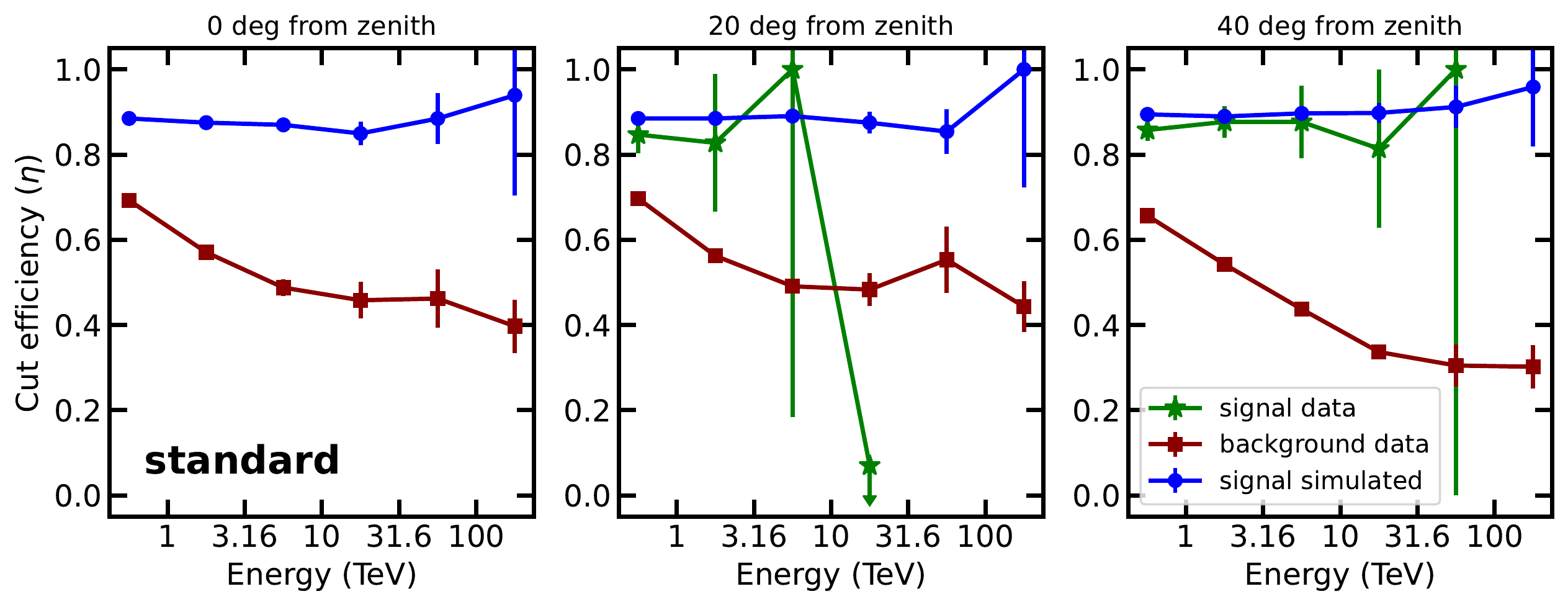}

	\caption{Fraction of events kept by the ABRIR cut applied after the \hess \textit{standard} cuts for simulated gamma-rays (blue dots), background data from off-runs (red squares) and events taken from a radius of 0.2\degree~from bright gamma-ray sources (green stars), in particular the blazar PKS 2155-304 at zenith angles of 20\degree~and the Crab Nebula for the 40\degree~zenith range. Note that since PKS 2155-304 is an extragalactic source, no gamma-rays are expected to arrive from it above a few TeV due to absorption on the extragalactic background light. When zero events survive the cut, the 68\% containment limit is drawn assuming Poissonian statistics.}
	\label{fig:rejection_power}       
\end{figure*}

\begin{figure*}
    \centering
	\includegraphics[width=0.95\textwidth]{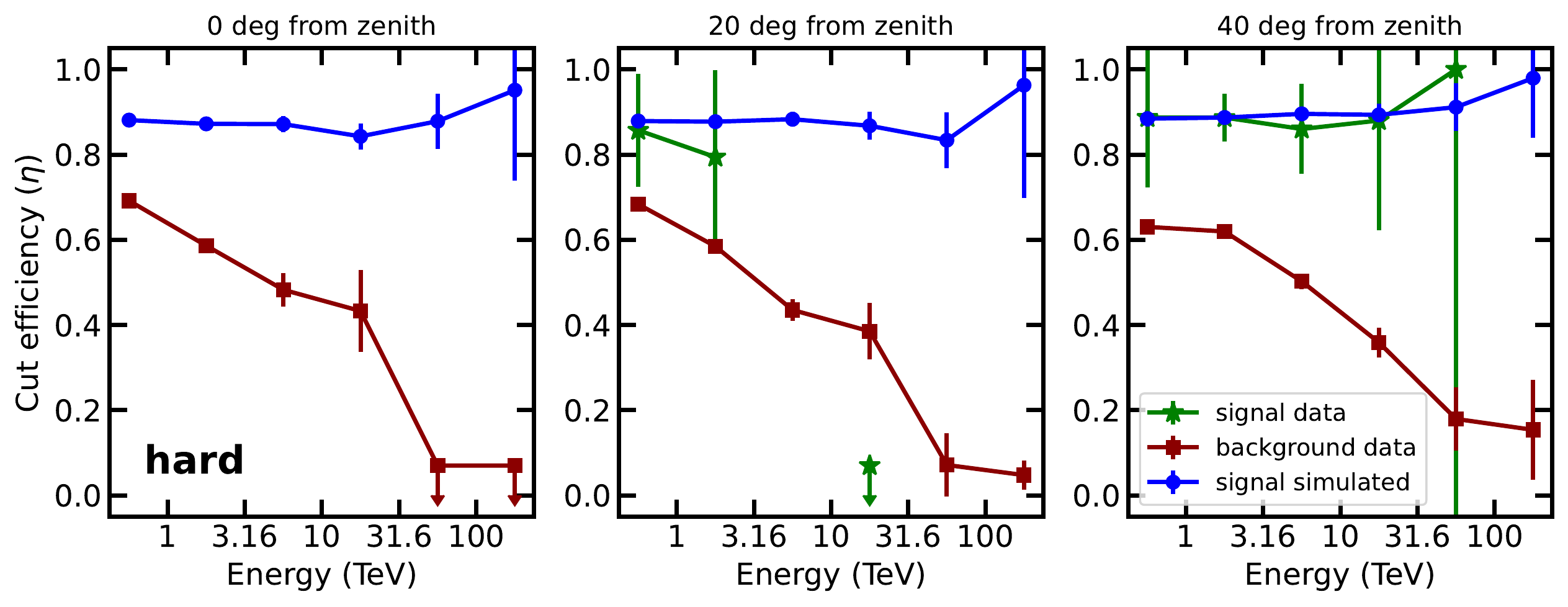}
	\caption{Fraction of events kept by the ABRIR cut applied after the \hess \textit{hard} cuts. Meaning of different panels and colors is the same as in~\ref{fig:rejection_power}. When zero events survive the cut, the 68\% containment limit is drawn assuming Poissonian statistics.}
	\label{fig:rejection_power_hard}       
\end{figure*}

We asses the performance of the algorithm by testing it on different types of events. Note that both the background rejection power and the gamma-ray efficiency presented in this section are a relative improvement on the value for the initial \hess \textit{standard} and \textit{hard} selection cuts. Those curves are shown in Figure~\ref{fig:rejection_power_stereo}.

\subsection{Background rejection power}
\label{subsec:rejection}
In order to test the performance of the algorithm on the cosmic-ray background, we use what is referred to as \textit{off-runs}, that is, observations in which the telescopes are pointed at fields without a known gamma-ray source. These off-runs are typically the result of dedicated observations of known empty fields, or also observations of extragalactic objects like dwarf spheroidal galaxies that did not yield a gamma-ray detection~\cite{dwarfsph}. These are not completely free of gamma-ray events, due to large-scale diffuse emission like the extragalactic gamma-ray background and also due to possible undetected faint gamma-ray sources. However, the relative fluxes of background cosmic-rays to these gamma-ray sources make it safe to neglect this contribution. It would also be possible to use simulations of proton-initiated showers in order to eliminate this contamination. However, due to uncertainties in the hadronic models used for high-energy particle interactions~\cite{harmdan,alvaro}, there are significant differences between the properties of measured background events and those that result from simulations. For this reason we decide to use real data only to characterize the background, as it is more realistic and not dependent on the agreement between data and simulations.

Figure~\ref{fig:example_off} shows two example rejected events from one of these off runs which pass the \textit{standard} \hess cuts. The right panel shows the ImPACT prediction associated with each event, whereas the left panel shows the actual recorded event image. For both cases, the conditions C1 and C2 are fulfilled.

Figures~\ref{fig:rejection_power} and~\ref{fig:rejection_power_hard} shows the ABRIR cut efficiency for different samples as a function of the event reconstructed energy for several observation zenith angles. Figure~\ref{fig:rejection_power} corresponds to the case where events surviving the \hess \textit{standard} cuts are given as input, whereas the input events for the curves in Figure~\ref{fig:rejection_power_hard} survive the \textit{hard} cuts. The efficiency shown in the figures is calculated simply as the ratio of the number of events before and after the application of ABRIR. As can be seen, the cut rejects a significant fraction of off-run events in both cases (red line) at all energies. Background rate reductions of up to a factor 2.5 are obtained for energies above a few TeV for the \textit{standard} cuts case, while when using only the higher quality events selected by the \textit{hard} cut as input, this improvement goes up to factors of 3 and even 4 for energies above tens of TeV.

\subsection{Gamma-ray efficiency}
\label{subsec:efficiency}
Besides rejecting a lot of suspected background events, it is important to keep a high fraction of the true gamma-ray events. Some gamma-ray events are expected to be flagged as rejected by the algorithm, and they belong to three different categories:
\begin{enumerate}
\sloppy
    \item \textbf{Gamma-ray events that contain muon light.} As shown by~\cite{ourpaper}, a small fraction of high-energy gamma-ray initiated showers will actually contain muons, which can be detected by the large telescope. These events make up an irreducible set of lost gamma-ray events associated with muon-tagging based rejection. 
    
    \item \textbf{Gamma-ray events with low-altitude electrons.} Since our approach does not identify muon light as such, the light from any other particle that would create a component in the image besides the gamma-ray shower could lead to an event being rejected. Camera images of gamma-ray showers can contain light from scattered electrons from the shower that emit close to the telescope dish. This can create additional image components separated from the main shower image. The likelihood of this effect drops rapidly as the shower core ground impact point moves away from the telescope position.

	\item \textbf{Gamma-ray events with unusually high NSB noise.} A fraction of gamma-ray events will contain noise from the NSB that is brighter than the threshold set by condition C2 from Section~\ref{subsec:algorithm}. This can happen in the simulations due to fluctuations in the NSB level, but it is also important in real data, especially since the NSB is not the same for all regions of the sky. This fraction is thus heavily dependent on the NSB conditions that were assumed when the algorithm's parameters were chosen, and also on the type of cleaning that is used to remove the NSB contribution to the images.
\end{enumerate}

We compute the gamma-ray efficiency of ABRIR by running it on simulated gamma-ray events, using the CORSIKA package~\cite{corsika} for shower and Cherenkov light simulation and the \textit{sim-telarray} package~\cite{sim-telarray} for the telescope response and camera simulation. In order to check the consistency of the result in real data, we also check with a sample of events which reconstructed direction falls within 0.2$\degree$ of the Crab Nebula and PKS 2155-304, which are known bright gamma-ray sources. Note that this sample is expected to contain some small fraction of cosmic-ray background, as cosmic-ray events are distributed roughly uniformly on the sky. However, due to the relatively high gamma-ray flux of these sources, the majority of events in this sample will be true gamma-rays, and thus can be used to check the performance on real data events. Note that this comparison is restricted to the zenith angles in which these sources can be observed by the \hess experiment and also by the energies reached by the sources themselves.

Figures~\ref{fig:rejection_power} and~\ref{fig:rejection_power_hard}  shows the ABRIR cut efficiency for the simulated gamma-ray sample (blue lines) as well as the events selected around known gamma-ray sources (green line). Note that the gamma-ray efficiency is mostly flat as a function of energy. This indicates that the group dominating the rejected gamma-rays are those with residual features caused by the NSB, since, unlike the expected number of muons, this does not depend on the event energy. The gamma-ray efficiency computed with real data is consistent with that found in simulations, which in turn confirms that the simulated NSB level is consistent with that encountered in a typical field.


\subsection{Application to a gamma-ray source: the Crab Nebula}
\label{subsec:data}
\begin{figure*}
  \includegraphics[width=0.95\textwidth]{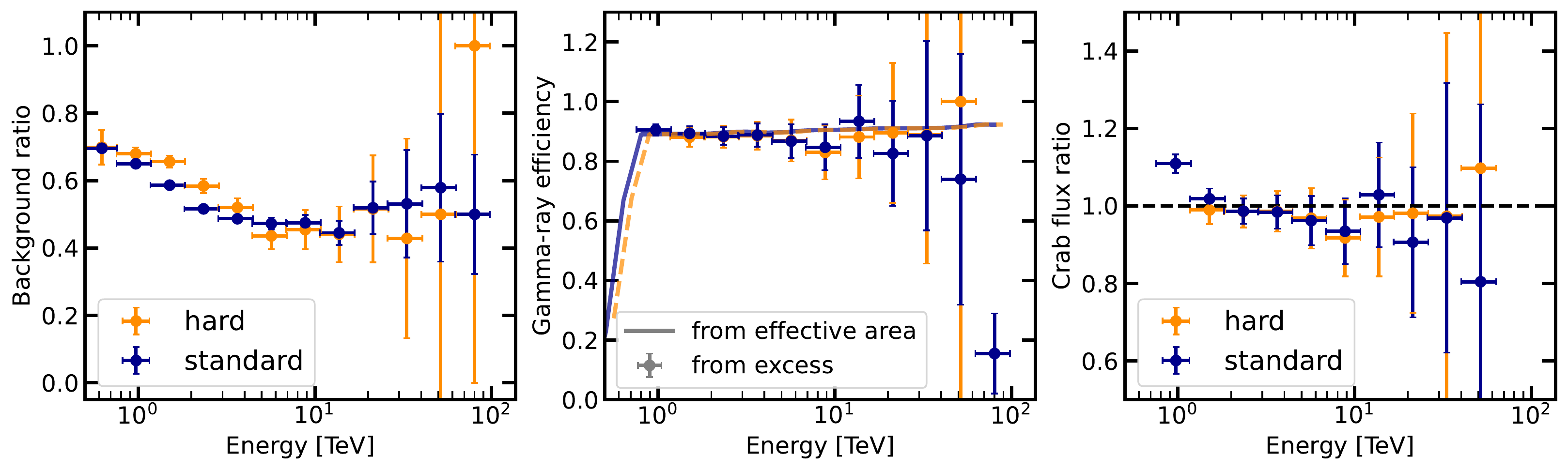}
\caption{Verification of the performance using data from the Crab Nebula. All ratios shown in this figure are computed as the quantity after applying the ABRIR cut divided by the same quantity before the cut. \textit{Left}: Ratio of background counts with and without the use of ABRIR for both the \textit{standard} (dark blue) and \textit{hard} (orange) cuts. \textit{Middle}: Gamma-ray efficiency computed as the ratio of the resulting effective area for the datasets with and  without ABRIR (solid lines) and as the ratio of the measured excess (data points), again for both sets of cuts using the same color scheme as in the left panel. \textit{Right}: Ratio of the measured flux from the Crab Nebula as a function of energy with and without the use of ABRIR for both sets of initial cuts.}
\label{fig:crab}       
\end{figure*}
In order to verify the expected improved performance, we perform an analysis of the same dataset from a real gamma-ray source with and without the use of ABRIR and compare the results. For this, we use a total of 30~h of observations of the Crab Nebula by the full array of the \hess telescopes. Note that since the goal of this paper is to demonstrate the relative performance of this analysis technique, only ratios of the relevant quantities will be shown. 

In both cases, we first perform a \textit{stereo} reconstruction of the data which excludes the large central telescope. This results in a list of events which, in the case without ABRIR are directly bundled with the relevant instrument response functions (IRFs) via the \textit{Gammapy}~\cite{gammapy} package, as described in~\cite{lars}. When applying ABRIR, an extra step is performed in which the algorithm is applied, the flagged events are removed from the event list and the effective areas are corrected by the gamma-ray efficiency computed from simulations. The new event lists and IRFs are bundled in the same way as before. 

Once both datasets are ready, we can extract and compare different quantities from them. First, we estimate the background level in a region of 0.3\degree~radius around the Crab location using the so-called \textit{reflected-region} background method (see~\cite{berge} for a detailed description). This choice of region size is relatively large for a standard \hess point-source analysis, but allows for increased statistics on the background measurement. The ratio of the measured background before and after applying ABRIR as a function of energy can be seen in the left panel of Figure~\ref{fig:crab}, for both sets of initial cuts. The resulting background efficiency is consistent with the one derived from the off runs shown in Figures~\ref{fig:rejection_power} and~\ref{fig:rejection_power_hard} within statistical errors. Using this measurement of the background, we can compute the excess counts in the source region, and compare it to the ratio of the effective areas. This is an additional verification that the performance on simulated gamma-rays is observed in real data. The result can be seen in the middle panel of Figure~\ref{fig:crab}. Finally we can compute the flux from the Crab nebula as a function of energy. The rightmost panel of Figure~\ref{fig:crab} shows the ratio of flux measured before and after applying ABRIR as a function of energy for both sets of cuts. As can be seen, they are consistent with a ratio of unity within statistical errors, with the exception of the lowest energy point in the \textit{standard} result. This indicates a possible mischaracterization of the effective area at the lowest energies. Given that effective areas were not produced from scratch for the ABRIR dataset, but rather derived from existing ones using an energy-dependent correction, it is possible that the region near threshold, where the effective area raises rapidly, is less accurately characterized. Dedicated IRFs or an increased threshold would solve this issue.

\subsection{Comparison to \textit{hybrid} reconstruction}
\label{subsec:hybrid}
The background rejection powers achieved with the combination of the baseline \textit{stereo} algorithms and ABRIR are superior than those obtained when data from the large telescope is included in the reconstruction from the beginning - i.e. \textit{hybrid} reconstruction. This can be seen when comparing the combination of the ABRIR efficiencies shown in Figures~\ref{fig:rejection_power} and~\ref{fig:rejection_power_hard} and the baseline \textit{stereo} efficiencies shown in Figure~\ref{fig:rejection_power_stereo} with those of the \textit{hybrid} analysis shown in Figure~\ref{fig:rejection_power_hybrid}. The reason for this lack of performance is that combining information from different telescope sizes at the reconstruction level is a non-trivial task, especially for the algorithms that rely on image shape parameters. Ongoing improvements to the \hess \textit{hybrid} chain have shown promising results and may achieve a performance comparable to that of ABRIR in the low energy range. However, at high energies
ABRIR will likely continue to provide improved performance beyond that achieved by a hybrid analysis based on the standard image parameters.

\section{Discussion}
\label{sec:discussion}
We have presented an algorithm that makes use of large-dish telescopes in an IACT array as a veto step in order to improve the background rejection. Its use improves the background rejection power of the baseline \hess \textit{stereo} reconstruction by a factor ranging between 2 and 4, depending on energy and the specifics of the initial cut. The combined efficiency of this extra cut with the baseline rejection power of the \textit{stereo} reconstruction can reach background rejection powers higher than 10$^4$ for high energies. Improved background rejection is crucial for the detection and characterization of extended and faint sources. Reducing the background rate also reduces the uncertainties associated with it, leading to improved precision at high energies. This improvement can be decisive in order to determine, for example, the presence of energy-dependent morphology.

The different performance at high energies depending on the chosen initial cut for the \textit{stereo} reconstruction is due to their different image amplitude thresholds. The \textit{standard} cuts keep events whose image has a total brightness of more than 60 p.e. in two of the four telescopes at least. The threshold for the \textit{hard} cuts is 200 p.e. The image of an event can have low brightness either because the primary particle had relatively low energy, or because the location of the shower core is far from the array. This means that some events that are below the 200 p.e. threshold will be reconstructed with high energies but large core distances. At large core distances, the number of muons that are detectable by the telescope will be reduced~\cite{ourpaper}, meaning that the power of the veto approach is reduced.

All results shown here are based on images for which the noise has been cleaned using the so-called \textit{tail-cuts} cleaning method, which requires each pixel to have an intensity exceeding a threshold $I_1$ and a neighbour exceeding a threshold $I_2$~\cite{2006A&A...457..899A}. In particular we have used $I_1=7$ p.e. and $I_2=4$ p.e., as required for the ImPACT algorithm~\cite{impact}. As mentioned in Section~\ref{sec:peformance}, remaining NSB noise seems to be the leading cause of loss of gamma-ray events when applying ABRIR. A different image cleaning approach, such as those based on the time information of the shower image~\cite{2013ICRC...33.3000S} could improve these results.

The presence of broken or unusable pixels in the camera, that is regions of the camera which do not take data, could in principle impact the ability to identify the presence of additional charge away from the shower~\cite{2006A&A...457..899A}. However, in the case of the \hess CT5 camera, the number of such pixels is usually very small. There are just 6 isolated permanently turned off pixels, and for certain sky regions, the presence of bright stars can result in one or two more pixels being turned off~\cite{Bi2022}. This leads to a very small effect given that the camera has 1764 pixels in total and muons are uniformly distributed across it.

We have shown here the performance of ABRIR when applied to the \hess array, which is made up of four middle-sized telescopes and one large, central telescope. The future Cherenkov Telescope Array (CTA) will be made up of telescopes of three sizes, the largest of which is smaller than the central \hess telescope yet still large enough to efficiently detect muons. The idea behind the algorithm will thus be applicable to the CTA array, although whether or not will it be a competitive technique will depend on the analysis approaches used for mixed-telescope types and sub-array selection of the observations. Large telescopes are typically seen as an asset only at the low energies due to their reduced threshold. However, as we show here, including large telescopes when observing targets for which the spectrum is expected to extend to high energies is worthwhile given the achievable improvements especially in background rejection.

\appendix
\setcounter{table}{0}
\renewcommand{\thetable}{A\arabic{table}}
\setcounter{figure}{0}
\renewcommand{\thefigure}{A\arabic{figure}}
\section{Efficiencies of the baseline \hess cuts}

\begin{figure*}
    \centering
	\includegraphics[width=0.95\textwidth]{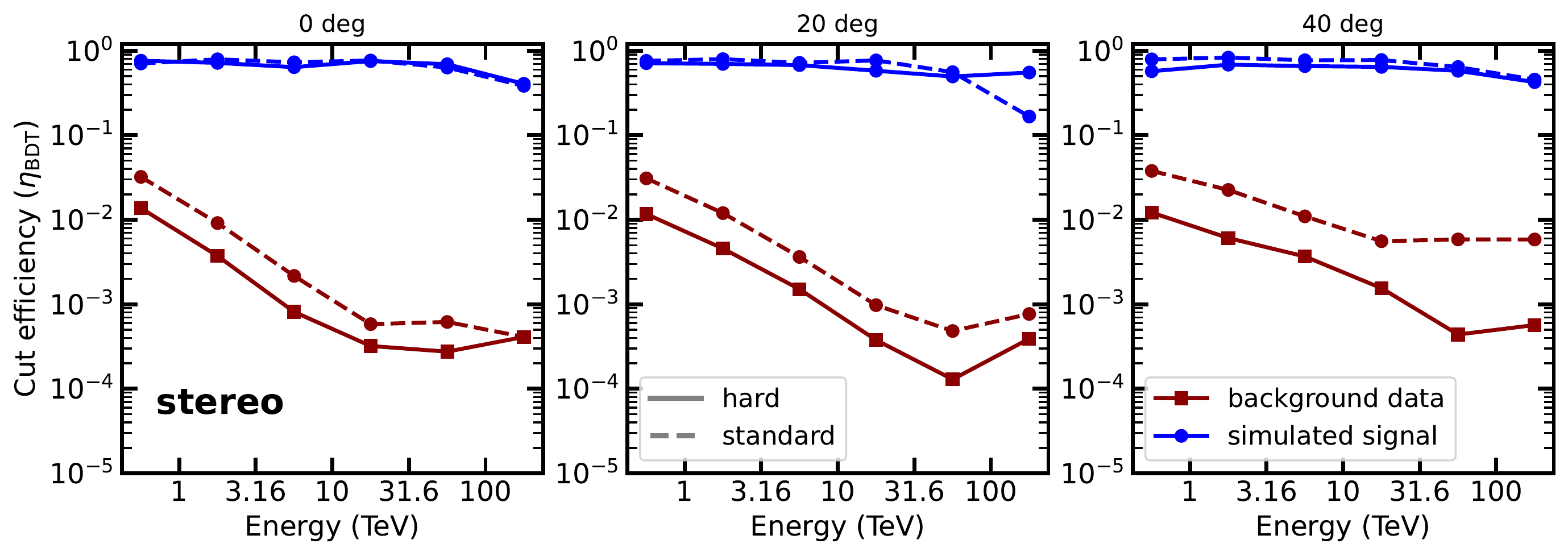}
	\caption{Cut efficiencies of the \textit{stereo} reconstruction for the \textit{standard} and \textit{hard} cuts.}
	\label{fig:rejection_power_stereo}       
\end{figure*}

\begin{figure*}
    \centering
	\includegraphics[width=0.95\textwidth]{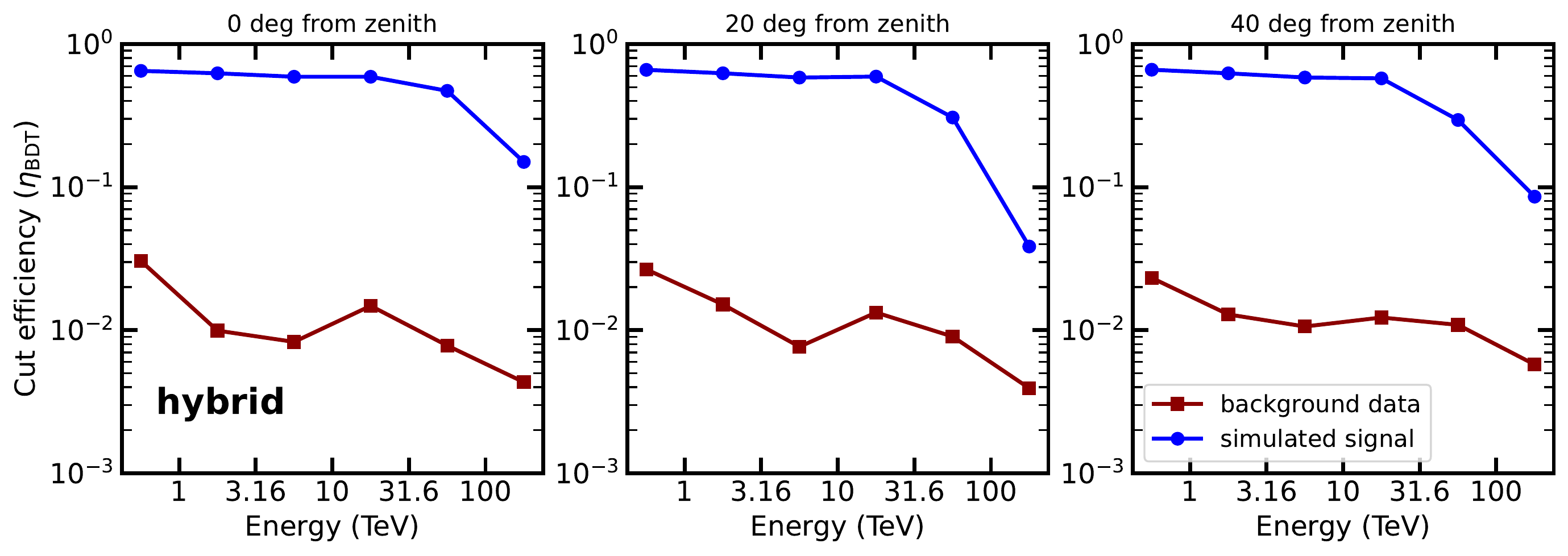}
	\caption{Cut efficiencies of the \textit{hybrid} reconstruction using the \textit{standard-hybrid} cuts.}
	\label{fig:rejection_power_hybrid}       
\end{figure*}

We computed the background and gamma-ray efficiencies by running the \hess analysis on the off run events and simulated gamma-rays with and without the baseline Boosted Decision Tree (BDT) cuts. The efficiency is then computed as the ratio of the events passing the BDT cut to the initial number of events. Note that this initial number comprises only events surviving the respective image amplitude threshold for each of the cut sets (60 p.e. for \textit{standard} cuts and 200 p.e. for the \textit{hard} cuts). The resulting efficiencies are shown for different energies and zenith ranges in Figures~\ref{fig:rejection_power_stereo} (\textit{stereo} reconstruction) and~\ref{fig:rejection_power_hybrid} (\textit{hybrid} reconstruction). Note that the cuts used in the \textit{hybrid} case, so-called \textit{standard-hybrid} cuts, are different from the \textit{stereo} ones in terms of thresholds, but are optimized for the same science case: a Crab Nebula-like source.

\section*{Declarations}

\textbf{Funding} This research was supported by the Max Planck Society. AMWM is supported by the Deutsche Forschungsgemeinschaft (DFG, German Research Foundation) – Project Number 452934793\\

\begin{acknowledgements}
The authors would like to thank the \hess Collaboration for allowing the use of \hess data and simulations in this publication, as well as providing useful discussions and input to the paper.
This work made use of \texttt{numpy}~\cite{numpy}, \texttt{scipy}~\cite{scipy}, \texttt{pandas}~\cite{pandas}, \texttt{matplotlib}~\cite{matplotlib}, \texttt{uproot}~\cite{uproot} and \texttt{ctapipe}~\cite{ctapipe}.
\end{acknowledgements}

\bibliographystyle{number_cite}       

\bibliography{refs}   

\end{document}